\documentstyle[psfig]{aa}

\begin{document}

\title{HST, VLT, and NTT imaging search for wide companions to bona-fide and candidate
brown dwarfs in the Cha I dark cloud \thanks{Based on observations obtained 
with the NASA/ESA Hubble Space Telescope in program GO 8716, 
ESO programs 64.I-0493, 66.C-0022, and 66.C-0310
as well as data taken from the public VLT data archive} }

\author{R. Neuh\"auser\inst{1}, W. Brandner\inst{2}, 
J. Alves\inst{2}, V. Joergens\inst{1}, F. Comer\'on\inst{2}}

\offprints{R. Neuh\"auser, rne@mpe.mpg.de }

\institute{MPI f\"ur extraterrestrische Physik, Giessenbachstra\ss e 1, D-85740 Garching, Germany
\and European Southern Observatory, Karl-Schwarzschild-Stra\ss e 2, D-85748 Garching, Germany
}

\titlerunning{Visual companion of brown dwarfs in Cha I}

\date{Received $<$date$>$ / Accepted $<$date$>$ } 

\abstract{We present results from a deep imaging search for companions around the
young bona-fide and candidate brown dwarfs Cha H$\alpha$ 1 to 12 in the Cha I dark cloud,
performed with HST WFPC2 (R, I, H$\alpha$), VLT FORS1 (VRI), and NTT SofI (JHK$_{\rm s}$).
We find 16 faint companion candidates around five primaries with separations between
1.5$^{\prime \prime}$ and 7$^{\prime \prime}$ and magnitudes in R \& I from 19 to 25 mag, 
i.e. up to 8 mag fainter than the primaries. While most of these companion candidates
are probably unrelated background objects, there is one promising
candidate, namely $1.5^{\prime \prime}$ SW off the M6-dwarf
Cha H$\alpha$~5. This candidate is 3.8 to 4.7 mag fainter than the primary and
its colors are consistent with an early- to mid-L spectral type.
Assuming the same distance (140 pc) and absorption (A$_{\rm I}$ = 0.47 mag)
as towards the primary, the companion candidate has
$\log$~(L$_{\rm bol}/$L$_{\odot}) = -3.0 \pm 0.3$. At the age of the primary
(1 to 5 Myrs), the faint object would have a mass of 3 to 15 Jupiter masses according
to Burrows et al. (1997) and Chabrier \& Baraffe (2000) models.
The probability for this companion candidate to be an unrelated fore- or background object
is $\le 0.7\%$, its colors are marginally consistent with a strongly reddened 
background K giant.
One other companion candidate has infrared colors consistent with an early T-dwarf.
In addition, we present indications for Cha H$\alpha$ 2 being a close ($\sim 0.2^{\prime \prime}$)
binary with both components very close to the sub-stellar limit.
Our detection limits are such that we should have detected all companions above
$\sim 1$~M$_{\rm jup}$ with separations $\ge 2^{\prime \prime}$ ($\ge 320$~AU) and
all above $\sim 5$~M$_{\rm jup}$ at $\ge 0.35^{\prime \prime}$ ($\ge 50$~AU).
}

\maketitle

\section{Introduction: Multiplicity of brown dwarfs}

Extrasolar planets were detected by indirect methods,
but no direct imaging was presented, yet.
Imaging of sub-stellar companions around young very low-mass
stars or brown dwarfs (rather than around normal solar-type stars) should
be less difficult, because young sub-stellar objects are self-luminous
due to ongoing contraction and very low-mass stars and brown dwarfs as
primaries are intrinsically faint. Hence, the problem of dynamic range (large 
magnitude difference at small angular separation) is not that severe.
Only a few brown dwarfs were detected directly as companions to normal
stars so far, mostly around M-dwarfs, the first of which was Gl 229 B
(Nakajima et al. 1995, Oppenheimer et al. 1995), 
and the youngest of which is TWA-5~B
(Lowrance et al. 1999, Neuh\"auser et al. 2000).

Brandner et al. (2000) undertook an HST and AO survey for
sub-stellar companions around T Tauri stars, which we now extend to
the M6- to M8-type objects Cha H$\alpha$ 1 to 12 in the Cha I dark cloud.
We obtained H$\alpha$ as well as R- and I-band images with the Hubble Space
Telescope (HST) in order to search for wide visual companions around them.
These observations are both deeper and have higher spatial resolution than any previous
images of these objects. We complement the HST data with archived VRI data from the
ESO 8.2m Very Large Telescope (VLT) and
new JHK$_{\rm s}$ data from the ESO 3.5m New Technology Telescope (NTT).

With deep imaging of bona-fide and candidate brown dwarfs as presented here,
one can, in principle, 
find binaries comprised of (i) two low-mass stars 
(as some of the candidate brown dwarfs in the sample could be very low-mass stars), 
(ii) one brown dwarf and one low-mass star, (iii) two brown dwarfs, (iv) one 
giant planet and one low-mass star, (v) or one giant planet and one brown dwarf.

So far, only little is known about multiplicity of brown dwarfs:
The first binary brown dwarf found was PPl 15, a spectroscopic binary made up 
of two brown dwarfs in the Pleiades (Basri \& Mart\'\i n 1999). Then, faint 
companion candidates were detected directly near a few brown dwarfs, namely
DENISJ1228.2 (Mart\'\i n et al. 1999), 2MASSWJ1146 (Koerner et al. 1999),
and 2MASSJ0850 (Reid et al. 2001).

Multiplicity parameters of brown dwarfs (like binary frequency, 
orbit characteristics, mass functions of primaries and secondaries, etc.) 
will shed light on their as yet uncertain formation mechanism: 
Do brown dwarf companions form like stellar companions or like planets 
in circumstellar disks ? Do they form by fragmentation or core growth ?
Do isolated brown dwarfs form in isolation or are they ejected from multiple systems ?
If brown dwarfs are ejected in the early accretion phase (Reipurth \& Clarke 2000), 
then one should not find any brown dwarfs as companions at ages $\ge 1$ Myrs.
If low-mass companions are ejected during the pre-main sequence phase
(c.f. Sterzik et al. 2001), then young brown dwarf companions should 
be more frequent than old brown dwarf companions,
which we can test with our sample of young objects in Cha I.

It is as yet unknown, whether brown dwarfs can have planets and, if so,
what their typical separations from the primary objects would be.
With several high-resolution spectra obtained with VLT/UVES,
Joergens \& Guenther (2001) found evidence for radial velocity
variations in some of the Cha H$\alpha$ objects, 
which could be due to giant planets or surface features.

In Sect. 2, we will present the properties of the targets.
In Sect. 3, details about the HST observations and data reduction
are given. The complementary VLT and NTT observations
are presented in Sect. 4. The background contamination is 
estimated in Sect. 5. Properties of the most promising companion 
candidate are discussed in Sect. 6. 
In Sect. 7, we present indications for one of the primaries being
a very close $\sim 0.2^{\prime \prime}$ binary pair.
We conclude with a brief discussion in Sect. 8.

\section{Our targets: Cha H$\alpha$ 1 to 12}

The Cha I dark cloud is a site of ongoing low- and intermediate mass star 
formation at $\sim 140$ to 160 pc. Most of the known members are pre-main
sequence T Tauri stars, but there is also one known intermediate-mass
Herbig A0e star (HD 97048) and several M6- to M8-type bona-fide and
candidate brown dwarfs. The latter objects (Cha H$\alpha$ 1 to 12) were found in 
deep infrared (IR), H$\alpha$, and X-ray surveys (Neuh\"auser \& Comer\'on 1998, 1999;
Comer\'on et al. 1999, 2000, henceforth C1999 \& C2000), e.g.
Cha~H$\alpha$~1 was the first X-ray detected brown dwarf.
With ground-based optical and near-IR data as well as mid-IR data
from ISO (C2000, Persi et al. 2000),
Natta \& Testi (2001) show that Cha H$\alpha$ 1, 2, and 9 have
IR excess emission indicative for circumstellar material like disks.
Also, Wilking et al. (1999) and Muench et al. (2001) found evidence
for circum{\em stellar} material around young brown dwarfs in $\rho$ Oph
and the Trapezium, respectively.

Based on their location in the H-R diagram and comparison with different
theoretical isochrones, Cha H$\alpha$ 1 to 12 are roughly co-eval,
namely $\sim 2$ Myrs according to Baraffe et al. (1998)
and $\sim 1$ to 10~Myrs according to Burrows et al. (1997).
The M7.5-M8 type objects Cha H$\alpha$ 1, 7, 10, and 11 are clearly 
located below the sub-stellar limit and, hence, were classified as
bona-fide brown dwarfs, while the other Cha H$\alpha$
objects are candidate brown dwarfs (C2000).
The brightest of them, Cha H$\alpha$ 1 to 8 and 12, are known to show Lithium
absorption and radial velocity consistent with membership to the Cha I
T association (Neuh\"auser \& Comer\'on 1999, Joergens \& Guenther 2001),
the others were not observed with high resolution, yet, because of their faintness.
Previously, none of the bona-fide and candidate brown dwarfs 
Cha H$\alpha$ 1 to 12 observed here were known to be multiple.

\section{HST observation and data reduction}

In program 8716, we obtained deep imaging with the 
HST Wide Field Planetary Camera No. 2 (WFPC2) filters F656N (H$\alpha$), 
F675W (R-band), and F814W (I-band). We performed the HST observations                                       
in the optical (rather than in the IR), because NICMOS was not available during
the last HST cycle. For our WFPC observations, the two reddest broad-band 
filters were used (R \& I, because substellar objects are red)
as well as the H$\alpha$ filter; the latter was used, because
all the primary objects Cha H$\alpha$ 1 to 12 
show H$\alpha$ emission, so that any additional faint object with H$\alpha$
emission would be a good candidate for a hithertoo unknown brown dwarf.

Faint companions to any of the targets would be sub-stellar, because all our
(as yet unresolved) targets are located either below or very close to the
sub-stellar limit, so that any much fainter,
i.e. lower-mass companion would certainly be sub-stellar.

Because of missing guide stars (all targets are located inside the dark
cloud Cha I), Cha H$\alpha$ 6 could not be observed.
All other eleven targets were observed.

Whenever possible, we placed several objects in the field-of-view of the 
WFPC chips to maximize the efficiency. Isolated objects were placed onto the PC 
chip for better angular resolution: 
The PC chip has a 0.0455$^{\prime \prime}$/pixel scale, 
the WF chips have a scale of 0.0996$^{\prime \prime}$/pixel.
In total, we observed in eight HST orbits in 2000 and 2001.
See Table 1 for the HST observation log.

\begin{table}
\caption{Observing log of HST WFPC2 observations (*)}
\begin{tabular}{llrl} \hline
Object & filter & exposure & JD \\ 
in Cha I  &     & [seconds] & date \\ \hline
H$\alpha$ 1 \& 7 & F656N (H$\alpha$)  & 2 $\times$ 500   & 2451952 \\
          & F675W (R)    & 2 $\times$ 100   & 12 Feb 2001 \\
          & F814W (I)    & 2 $\times$ 100   & \\ \hline
H$\alpha$ 2 \& 9 & F656N (H$\alpha$)  & 2 $\times$ 500   & 2451842 \\
          & F675W (R)    & 2 $\times$ 100   & 25 Oct 2000  \\
          & F814W (I)    & 2 $\times$ 100   (**) & \\ \hline
H$\alpha$ 3      & F656N (H$\alpha$ ) & 2 $\times$ 700   & 2451733 \\
          & F675W (R)    & 300 + 80  & 8 Jul 2000 \\
          & F814W (I)    & 160 + 40  & \\ \hline
H$\alpha$ 4,10,11 & F656N (H$\alpha$) & 4 $\times$ 500   & 2451783 \\
\& 8 (***) & F675W (R)    & 2$\times$(300+100) & 27 Aug 2000 \\
          & F814W (I)    & 2$\times$(300+100) & \\ \hline
H$\alpha$ 5      & F656N (H$\alpha$)  & 2 $\times$ 700   & 2451778 \\
          & F675W (R)    & 300 + 80  & 22 Aug 2000 \\
          & F814W (I)    & 160 + 40  & \\ \hline
H$\alpha$ 8      & F656N (H$\alpha$)  & 2 $\times$ 700   & 2451752 \\
          & F675W (R)    & 300 + 80  & 27 Jul 2000 \\
          & F814W (I)    & 160 + 40  & \\ \hline
H$\alpha$ 12     & F656N (H$\alpha$)  & 2$\times$(260+70) & 2451914 \\
          & F675W (R)    & 2$\times$(10+40) & 05 Jan 2001 \\
          & F814W (I)    & 2$\times$(10+40) & \\ \hline 
\end{tabular}
\vspace{+.2cm}
\small{Remarks: 
\vspace{-.2cm}
(*) There are no other (archived or scheduled) HST observations
with any of the targets in the field-of-view (as of Sept. 2001).
(**) The observations in the I filter failed due to technical 
problems, hence no data in I. (***) Cha H$\alpha$ 8 and a few of its 
companion candidates were again in the field of these observations,
namely at the edge of chip WF2.}
\end{table}

Data reduction was performed with the IRAF\footnote{IRAF is distributed by
the National Optical Astronomy Observatories, which is operated by the
Association of Universities for Research in Astronomy, Inc. (AURA) under
cooperative agreement with the National Science Foundation.}
package {\em stsdas} written for HST data. 
All exposures (for each field and each filter) were split into two
to four to facilitate for cosmic ray rejection. After rejection of 
cosmic rays with {\em crrej}, we co-added the data. 
Mostly, the background was very small ($\sim 1$ count),
because the targets are in the dark cloud Cha I, while the background
was slightly higher near the bright (V=8.5 mag) A0e star HD 97048.

Source detection was done by visual inspection. A faint, non-extended 
object near any of the targets, i.e. within one or a few arc sec, 
can in principle be either a real companion or an unrelated fore- or background
object. Since we have no information, yet, on the spectra or proper motions
of these faint objects, all of them have to be regarded as companion candidates,
unless companionship can already be ruled out from color or extension.
For the maximum separation between a primary target and a potential companion, we 
allow for (more or less arbitrarily) 1000 AU, which is $\sim 7^{\prime \prime}$ 
at the distance of Cha I ($\sim 140$ to 160 pc). 
Hence, in Table 2, we list only those faint objects,
which are found within 7$^{\prime \prime}$ around the Cha H$\alpha$ targets, while
other faint objects are regarded as unrelated objects most likely located in the 
background of Cha I. The observed background population is used in section 5
for a rough estimation as to how likely the companion candidates are true companions 
rather than unrelated objects.

In Figs. 1 and 2, we show the HST I-band images of those Cha H$\alpha$ objects,
where several companion candidates were detected within 7$^{\prime \prime}$,
namely Cha H$\alpha$ 4, 10, \& 11 (Fig. 1) and Cha H$\alpha$ 8 (Fig. 2), 
while Cha H$\alpha$ 5 is shown in Figs. 8 to 10.
The faint companion candidates are labeled with their cc numbers given in Table 2
({\em cc} for {\em c}ompanion {\em c}andidate).
The companion candidates Cha H$\alpha$ 8/cc 1 \& 2 are detected only in the
observation which mainly aimed at Cha H$\alpha$ 4, 10, and 11, where Cha H$\alpha$
8 was located at the edge of chip WF2. Because one of the main targets, Cha H$\alpha$ 6,
was not observable due to the lack of guide stars, we assigned the extra observing
time (one more orbit) for the observation with Cha H$\alpha$ 4, 10, \& 11 on the chip WF4,
because of the large number of primaries observable simultaneously. 
Only due to the increased exposure time (800 sec compared to 200 sec 
in the I-band), the faint companion candidates Cha H$\alpha$ 8/cc 1 \& 2 were 
detected. They are detected only in I, see Fig. 2. Due to the longer exposure time in
this observation, Cha H$\alpha$ 4 got saturated in I.

\begin{figure}
\caption{HST I-band image of Cha H$\alpha$ 4, 10, and 11 and surrounding field,
exposure time is 800 sec, obtained on 27 Aug 2000.
The circles around the three main targets each have 7$^{\prime \prime}$ radii,
located on the WF4 chip. {\bf See A\&A paper for figure.}}
\end{figure}

\begin{figure*}
\vbox{\psfig{figure=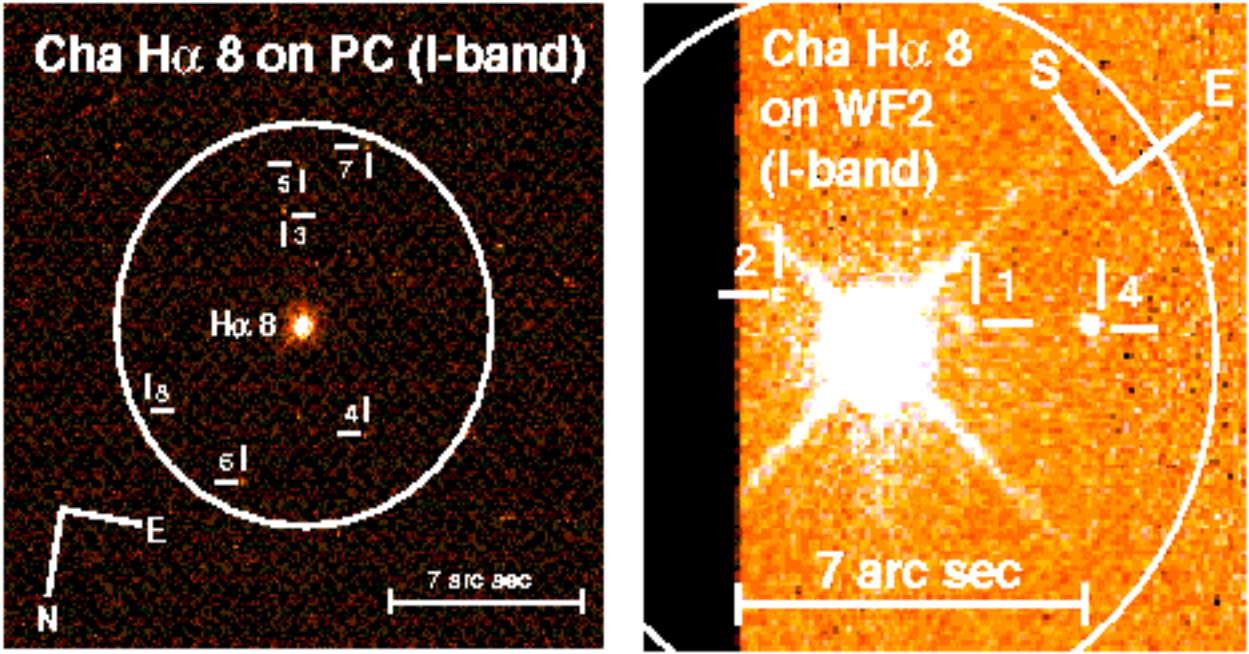,width=17cm,height=9cm}}
\caption{HST I-band images of Cha H$\alpha$ 8 and surrounding field:
The left image shows the 200 sec integration on the PC chip (obtained on 27 July 2000),
the right panel shows the 800 sec integration (27 Aug 2000), where the target is
located at the edge of the chip WF2, where two more faint companion
candidates were detected. The circles have 7$^{\prime \prime}$ radii.}
\end{figure*}

\begin{table}
\caption{List of all wide companion candidates found in the HST imaging data.
The object designations contain the J2000.0 positions.
In the remainder of the paper, we use the {\it other names},
where the designation {\em cc} means {\em c}ompanion {\em c}andidate.
The offsets (typically better than $\pm 0.1$$^{\prime \prime}$) 
of the companion candidates 
(sorted by increasing separations up to 7$^{\prime \prime}$, the upper limit) 
are given with respect to the primary.
Spectral types for the primaries are from C2000.}
\begin{tabular}{llcll} \hline
Object & other name & HST & \multicolumn{2}{c}{separation [$^{\prime \prime}$]} \\
designation & & chip & $\Delta \alpha$ & $\Delta \delta$ \\ \hline
110717.0-773454 & Cha H$\alpha$ 1    & WF4 & primary & M7.5 \\ \hline
110743.0-773359 & Cha H$\alpha$ 2    & WF4 & primary & M6.5 \\ \hline
110752.9-773656 & Cha H$\alpha$ 3    & PC  & primary & M7 \\ \hline
110819.6-773917 & Cha H$\alpha$ 4    & WF4 & primary & M6 \\
110819.7-773918 & H$\alpha$ 4/cc 1 & WF4 & 2.0 E & 1.3 S \\ 
110819.5-773920 & H$\alpha$ 4/cc 2 & WF4 & 1.4 W & 3.3 S \\
110819.8-773918 & H$\alpha$ 4/cc 3 & WF4 & 3.5 E & 1.3 N \\ \hline
110825.6-774146 & Cha H$\alpha$ 5  & PC & primary & M6 \\
110825.5-774147 & H$\alpha$ 5/cc 1  & PC & 1.0 W & 1.1 S \\ \hline
110840.2-773417 & Cha H$\alpha$ 6 & \multicolumn{2}{l}{not observed} & M7 \\ \hline
110738.4-773530 & Cha H$\alpha$ 7 & WF2 & primary & M8 \\ \hline
110747.8-774008 & Cha H$\alpha$ 8 & PC & primary & M6.5 \\
110747.9-774007 & H$\alpha$ 8/cc 1 & WF2 & 1.4 E & 1.3 N \\
110747.8-774010 & H$\alpha$ 8/cc 2 & WF2 & 0.4 W & 2.3 S \\ 
110747.7-774012 & H$\alpha$ 8/cc 3 & PC  & 1.3 W & 4.0 S \\
110748.0-774004 & H$\alpha$ 8/cc 4 & PC  & 2.9 E & 3.5 N \\
110747.7-774014 & H$\alpha$ 8/cc 5 & PC  & 0.8 W & 5.8 S \\
110747.6-774003 & H$\alpha$ 8/cc 6 & PC  & 3.0 W & 5.3 N \\
110747.9-774015 & H$\alpha$ 8/cc 7 & PC  & 1.7 E & 6.7 S \\
110747.4-774006 & H$\alpha$ 8/cc 8 & PC  & 6.6 W & 2.4 N \\ \hline
110719.2-773252 & Cha H$\alpha$ 9 & WF2  & primary & M6 \\ \hline
110825.6-773930 & Cha H$\alpha$ 10 & WF4 & primary & M7.5 \\ 
110825.8-773930 & H$\alpha$ 10/cc 1 & WF4 & 3.4 E & 0.3 S \\
110825.7-773927 & H$\alpha$ 10/cc 2 & WF4 & 2.2 E & 3.2 N \\
110825.3-773932 & H$\alpha$ 10/cc 3 & WF4 & 5.2 W & 1.7 S \\ \hline
110830.8-773919 & Cha H$\alpha$ 11 & WF4 & primary & M8 \\ 
110830.6-773915 & H$\alpha$ 11/cc 1 & WF4 & 2.7 W & 4.0 N \\ \hline
110637.4-774307 & Cha H$\alpha$ 12 & WF4 & primary & M7 \\ \hline
\end{tabular}
\end{table}

Aperture photometry was obtained for all objects (detected by visual inspection)
with a 0.5$^{\prime \prime}$ radius with a source-free background 
field annulus located around 
the measured object. The counts were transformed to magnitudes $m$ using 
\begin{equation}
m [mag] = -2.5 \cdot \log (DN / exp ) + ZP - 0.1
\end{equation}
with the number of counts {\it DN}, the exposure time {\it exp} in seconds, and 
the zero point {\it ZP}; the subtraction of 0.1 mag is to correct 
from the 0.5$^{\prime \prime}$ radius to the infinite aperture; all according to the HST
data reduction manual.\footnote{http://www.stsci.edu/documents/dhb/web}

Magnitudes obtained with Eq. (1) have still to be corrected 
for the charge-transfer-efficiency (CTE) loss.
This was done according to Whitmore et al. (1999) with the number of counts,
the observational date, the background level, and the (x,y) pixel position of the
source peak as inputs. The CTE corrections are especially important for weak sources 
such as the brown dwarf companion candidates expected in our program.
The final values are given in Table 3.

\begin{table*}
\caption{Ground-based V,R$_{\rm C}$,I$_{\rm C}$ from VLT/FORS1,
space-based R$_{\rm V}$,I$_{\rm V}$,m(H$\alpha$) from HST/WFPC2,
and ground-based J,H,K$_{\rm s}$ from NTT/SofI: 
The ground-based VRI data for the primary objects are from C2000,
the JHK$_{\rm s}$ for Cha H$\alpha$ 1, 2, 3, 7, and 9 are also from C2000, 
but for all other primaries and all companion candidates
from our NTT data, the ground-based VRI magnitudes for the companion 
candidates were determined by us from the C2000 VLT/FORS1 images.
Typical precision of the magnitudes are $\pm 0.05$ for the bright primary objects 
and $\pm 0.1$ for the companion candidates, but $\pm 0.2$ mag or worse for objects fainter 
than 23 mag as well as Cha H$\alpha$ 5/cc 1 because of its small separation (see Sect. 6); 
n/o for not observed; see text for determination of lower magnitude limits.
We list the observed range in H$\alpha$ equivalent width W$_{\lambda}$ 
from C1999, Neuh\"auser \& Comer\'on 1999, and C2000, H$\alpha$ always in emission.}
\begin{tabular}{l|r|rr|rr|rrr|cc} \hline
Object & V & R$_{\rm C}$ & m(F675W) & I$_{\rm C}$ & m(F814W) & J & H & K$_{\rm s}$ & m(F656N) & W$_{\lambda}$(H$\alpha$)\\ 
 & (VLT) & (VLT) & (HST R$_{\rm V}$) & (VLT) & (HST I$_{\rm V}$) & & & & (HST H$\alpha$) & [\AA ] \\ \hline
Cha H$\alpha$ 1   & 21.0     & 18.7     & 18.96    & 16.17    & 16.16& 13.55 & 12.78 & 12.28 & 17.72 & 35-99 \\ \hline
Cha H$\alpha$ 2   & 19.8     & 17.60    & 17.42    & 15.08    & n/o  & 12.59 & 11.43 & 11.15 & 16.76 & 32-71 \\ \hline
Cha H$\alpha$ 3   & 19.51    & 17.38    & 17.27    & 14.89    & 14.97& 12.46 & 11.64 & 11.11 & 16.81 & 5-14 \\ \hline
Cha H$\alpha$ 4   & 18.52    & 16.70    & 16.68    & 14.34    & sat. & 12.19 & 11.38 & 11.09 & 16.69 & 5-11 \\
H$\alpha$ 4/cc 1  &$\ge 22.5$&$\ge 22.2$&$\ge 23.6$&$\ge 19.0$& 23.0 &$\ge 18.4$&$\ge 18.1$&$\ge 17.3$&       & \\ 
H$\alpha$ 4/cc 2  &$\ge 22.8$&$\ge 22.3$& 24.1     &$\ge 20.7$& 22.2 & 18.5  & 17.4  & 17.3  &       & \\
H$\alpha$ 4/cc 3  & 23.0     & 21.6     & 21.7     & 20.4     & 20.2 & 17.7  & 16.7  & 16.2  &       & \\ \hline
Cha H$\alpha$ 5   & 19.18    & 17.14    & 17.03    & 14.68    & 14.67& 12.14 & 11.21 & 10.56 & 16.62 & 8-11 \\
H$\alpha$ 5/cc 1  & 23.1:    & 21.1:    & 21.22    & 19.0:    & 19.20& 16.8: & 15.0: & 14.4: &       & \\ \hline
Cha H$\alpha$ 6   & 19.75    & 17.60    & n/o      & 15.13    & n/o  & 12.43 & 11.61 & 11.09 & n/o   & 59-76 \\ \hline
Cha H$\alpha$ 7   & 22.2     & 19.5     & 19.69    & 16.86    & 16.78& 13.89 & 13.00 & 12.51 & 19.02 & 35-45 \\ \hline
Cha H$\alpha$ 8   & 20.1     & 17.96    & 17.84    & 15.47    & 15.50& 12.9  & 12.0  & 11.4  & 17.46 & 8-9 \\ 
H$\alpha$ 8/cc  1 &$\ge 23.2$&$\ge 21.5$&$\ge 23.5$&$\ge 19.3$& 23.4 &$\ge 17.5$&$\ge 16.6$&$\ge 14.7$&       & \\
H$\alpha$ 8/cc  2 &$\ge 23.6$&$\ge 22.5$&$\ge 23.5$&$\ge 20.4$& 23.6 &$\ge 17.0$&$\ge 18.1$&$\ge 16.6$&       & \\
H$\alpha$ 8/cc  3 &$\ge 23.3$& 23.2     & 22.7     & 21.3     & 21.3 & 18.8  & 16.7  & 15.8  &       & \\
H$\alpha$ 8/cc  4 &$\ge 23.3$& 22.9     & 22.9     & 21.5     & 21.7 & 19.9  & 17.8  & 17.0  &       & \\
H$\alpha$ 8/cc  5 &$\ge 23.5$&$\ge 23.2$&$\ge 24.8$&$\ge 22.0$& 21.7 &$\ge 20.3$&$\ge 20.1$&$\ge 18.3$&       & \\
H$\alpha$ 8/cc  6 &$\ge 23.4$&$\ge 23.1$& 23.1     &$\ge 21.9$& 22.6 &$\ge 20.6$&$\ge 20.0$&$\ge 19.3$&       & \\
H$\alpha$ 8/cc  7 &$\ge 23.5$& 22.9     & 22.5     & 21.0     & 21.0 & 18.4  & 16.8  & 15.4  &       & \\
H$\alpha$ 8/cc  8 &$\ge 23.4$&$\ge 23.1$&$\ge 24.9$&$\ge 21.6$& 22.6 &$\ge 20.6$&$\ge 19.9$&$\ge 18.9$&       & \\ \hline
Cha H$\alpha$ 9   & 23.1     & 20.1     & 20.05    & 17.34    & n/o  & 13.92 & 12.59 & 11.82 & 20.02 & 16 \\ \hline
Cha H$\alpha$ 10  & 21.6     & 19.4     & 19.35    & 16.90    & 16.79& 14.41 & 13.68 & 13.30 & 19.21 & 9 \\ 
H$\alpha$ 10/cc 1 &$\ge 23.1$&$\ge 23.1$& 23.3     & 21.7     & 21.8 & 19.0  & 17.6  & 17.1  &       & \\
H$\alpha$ 10/cc 2 &$\ge 23.5$&$\ge 23.0$& 24.9     &$\ge 21.2$& 22.9 &$\ge 18.5$&$\ge 20.3$&$\ge 19.9$&       & \\ 
H$\alpha$ 10/cc 3 &$\ge 23.3$& 23.0     & 23.0     & 21.4     & 21.5 & 19.6  & 18.5  & 17.9  &       & \\ \hline
Cha H$\alpha$ 11  & 21.9     & 19.9     & 19.69    & 17.35    & 17.41& 14.72 & 14.03 & 13.60 & 18.95 & 23 \\ 
H$\alpha$ 11/cc 1 &$\ge 22.9$&$\ge 22.9$& 23.4     & 22.2     & 22.2 & 21.1  & 19.3  & 18.7  &       & \\ \hline
Cha H$\alpha$ 12  & 20.6     & 18.3     & 18.34    & 15.58    & 15.72& n/o   & n/o   & n/o   & 18.03 & 20 \\ \hline
\end{tabular}
\end{table*}

A comparison of the HST H$\alpha$ magnitudes with previously known gound-based VLT
H$\alpha$ equivelent widths (C2000) shows a clear trend (see Fig. 3), with a scatter 
indicating strong H$\alpha$ variability, not unexpected for such young objects.
In Fig. 3, we plot the ground based H$\alpha$ equivalent width versus the
difference between HST H$\alpha$ and R-band magnitude, to really compare the
H$\alpha$ line flux above the continuum with the equivalent width.

\begin{figure}
\vbox{\psfig{figure=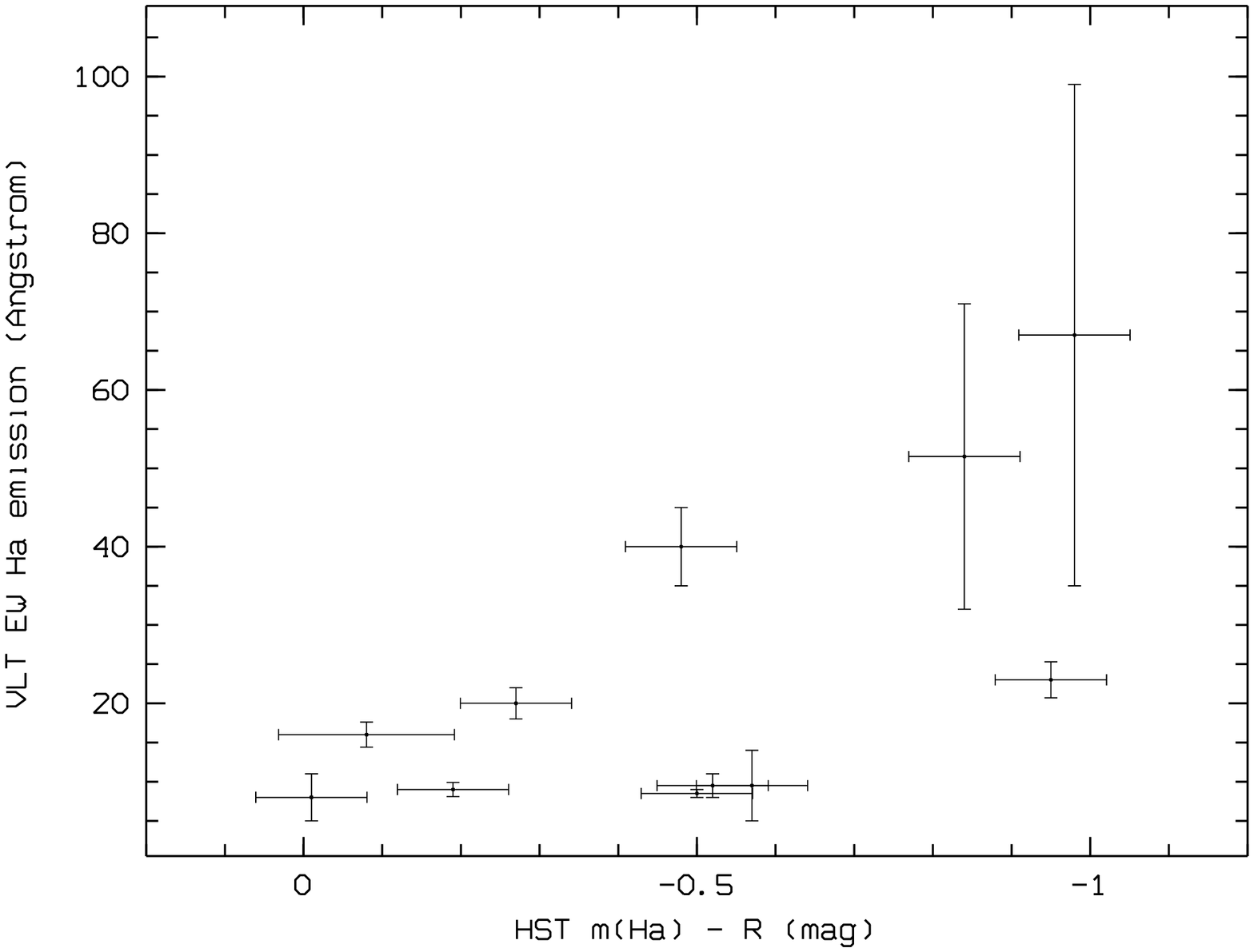,width=10cm,height=7cm,angle=270}}
\caption{Ground-based H$\alpha$ equivalent width (with its observed range due 
to variability or an assumed 10\% error bar) versus HST H$\alpha$ measurement
(plotted as difference between HST Vega H$\alpha$ magnitude and HST Vega R-band 
magnitude, so that we plot the line strength above continuum as for the equivalent width)
for the Cha H$\alpha$ primaries (none of the companion candidates were detected 
in H$\alpha$), data from Table 3. There is a trend for a correlation:
The brighter the HST H$\alpha$ magnitude (after subtraction of R), the
larger the ground-based H$\alpha$ equivalent width, as expected,
with some scatter most certainly due to variability, as expected.}
\end{figure}

%
%
%

\section{Ground-based imaging with VLT and NTT}

We also used archived VRI data\footnote{obtained
after commissioning of FORS1 before the start of normal operations; 
the data can be retrieved from the public VLT data archive; 
see C2000 for details, where the same data were used to
estimate the VRI values of Cha H$\alpha$~1 to 12}
obtained with the FOcal Reducer/low dispersion Spectrograph No. 1 (FORS1)
at the ESO 8.2m telescope Antu, Unit Telescope 1 (UT1) of the VLT,
with a scale of $0.2 ^{\prime \prime}$/pixel, during sub-arc sec conditions.
See Table 4 for the observations log.

We re-reduced the ground-based images.
Some companion candidates were also detected in the VLT VRI images,
magnitudes are given in Table 3.
Comparison of the HST R \& I magnitudes (Vega system) for all the
Cha H$\alpha$~1 to 12 objects with the previously known ground-based R \& I
magnitudes (Cousins bands) from C2000 shows that they agree well ($\pm 0.1$ mag).

\begin{table}
\caption{Observations log of ground-based images}
\vspace{-.3cm}
\begin{tabular}{llcrcc} \hline
Instr. & Cha H$\alpha$ & \hspace{-.2cm}{filter} & exp. & date & \hspace{-.3cm} FWHM \\ \hline
FORS1 & 4,5,8,10,11 & V      & 60s    & 26 Jan 1999 & 0.78" \\
FORS1 & 4,5,8,10,11 & R      & 60s    & 26 Jan 1999 & 0.87" \\
FORS1 & 4,5,8,10,11 & I      & 60s    & 26 Jan 1999 & 0.80" \\ \hline
SofI  & 4,5,8,10,11 & H           & 200s  & 14 Mar 2000 & 0.68" \\
SofI  & 4,5,8,10,11 & K$_{\rm s}$ & 200s  & 14 Mar 2000 & 0.58" \\ \hline
SofI  & 4,10,11     & J           & 600s  &  3 Mar 2001 & 0.60" \\
SofI  & 4,10,11     & H           & 1200s &  3 Mar 2001 & 0.58" \\
SofI  & 4,10,11     & K$_{\rm s}$ & 5400s &  3 Mar 2001 & 0.73" \\ \hline
SofI  & 5           & J           & 600s  &  7 Mar 2001 & 0.74" \\
SofI  & 5           & K$_{\rm s}$ & 5400s &  4 Mar 2001 & 0.58" \\ \hline
SofI  & 8           & J           & 600s  &  3 Mar 2001 & 0.64" \\
SofI  & 8           & H           & 1200s &  3 Mar 2001 & 0.72" \\ \hline
\end{tabular}
\end{table}

IR images in JHK$_{\rm s}$ were obtained using the IR imager Son of Isaac (SofI) 
at the ESO 3.5m NTT on La Silla with $\sim 0.6$ to 0.7$^{\prime \prime}$ seeing.
Images of three fields were taken in March 2001 with the 
{\em small SofI field} to achieve the highest angular
resolution possible ($0.147 ^{\prime \prime}$/pixel) 
with 2 sec individual exposure per frame.
Two more images (H \& K$_{\rm s}$) were already taken in March 2000 
with the {\em large SofI field} ($0.29 ^{\prime \prime}$/pixel), 
so that several Cha H$\alpha$ primaries were in the field-of-view, 
with 3 sec individual exposure per frame. 
Darks, flats, and standards were taken in the same nights,
and we performed standard data reduction with 
{\em eclipse}\footnote{see www.eso.org/projects/aot/eclipse/}
and {\em MIDAS}.\footnote{see www.eso.org/projects/esomidas/}
See Table 4 for the observations log and Table 3 for the JHK$_{\rm s}$ 
magnitudes. Our JHK$_{\rm s}$ data for the primaries (Table 3) agree
well with those given in C2000, who had obtained their data
with IRAC2 at the ESO/MPG 2.2m.

The new IJK$_{\rm s}$ data listed in Table 3 for the primaries also agree
well with the DENIS magnitudes listed in Neuh\"auser \& Comer\'on (1999)
for Cha H$\alpha$ 1 to 8. The H-band magnitude of Cha H$\alpha$ 8 (Table 3) 
was not known before, as this object was not observed in the IR by C2000 
and as the H-band filter is not used for DENIS.
The J \& K$_{\rm s}$ magnitudes for Cha H$\alpha$ 8 given
in Table 3 (from the March 2000 image) agree with the DENIS
magnitudes given in Neuh\"auser \& Comer\'on (1999).

Most of our ground-based VRIJHK$_{\rm s}$ magnitudes were determined by 
normal aperture photometry, with the exception for the VRIJ-band images 
for the very close companion candidate near Cha H$\alpha$ 5, 
which is discussed in detail in Sect. 6.

Lower limits to the magnitudes for undetected objects are derived from mean background
intensity (and its variation) from $\pm 4$ pixels (for HST images) 
and $\pm 3$ pixels (for ground-based images) around the pixel, where the 
peak of the companion candidate should be located according to its position.
The limits given correspond to an intensity of $3 \sigma$ above the mean background.
No additonal companion candidates (within 7$^{\prime \prime}$) were 
detected in the ground-based images.

\section{Real companions or background objects~?}

A first check, whether or not a companion candidate may be truely bound 
(hence at the same distance and age as primary),
i.e. whether or not it is as cool as expected from the magnitude difference,
can be done with the optical and IR colors. For several companion candidates, 
one can find a solution for the observed magnitudes and colors for either 
a moderately reddened L- or T-type companion or, alternatively,
for a highly reddened background star.
In the next section, we go through such an estimate for the
companion candidate near Cha H$\alpha$ 5.

Figs. 4 and 5 show that most companion candidates have colors similar
to background giants, while some of them have colors which are (also)
consistent with L-dwarfs.
Of those, Cha H$\alpha$ 5/cc 1 has the smallest separation and is
therefore the best companion candidate (see next section for a detailed
discussion of this object).
Given their faint magnitudes, all these companion candidates would
have masses around or below the deuterium burning limit.
For such low masses, they should be very faint and red in the optical.
However, with the possible exception of 
Cha H$\alpha$ 5/cc 1 and Cha H$\alpha$ 4/cc 2,
they are all too bright in R \& I given their JHK$_{\rm s}$ values,
so that they are probably reddened background giants or extragalactic.
Cha H$\alpha$ 8/cc 7 is located in the area of IR excess (Fig. 5),
possibly indicating circumstellar material.

Another possibly interesting object is Cha H$\alpha$ 4/cc 2 with
J$-$H = $1.10 \pm 0.14$ and H$-$K$_{\rm s} = 0.10 \pm 0.14$ mag,
the bluest in H$-$K$_{\rm s}$ in Fig. 5.
The T3-dwarf SDSSpJ102109.69-030420.1 has similar (2MASS) colors 
according to Burgasser et al. (2001), namely J$-$H = $0.93 \pm 0.15$
and H$-$K$_{\rm s} = 0.23 \pm 0.20$ mag\footnote{but somewhat different 
colors according to Leggett et al. (2002), namely J$-$H$\simeq 0.47$
and H$-$K$\simeq 0.15$ mag}.
If this object would really be an early T-dwarf companion,
it would definitely be a planetary mass-object.
Its separation, however, is 3.6$^{\prime \prime}$, 
i.e. $\sim 500$ AU at 140 pc, larger than expected for planets.

Whether a certain companion candidate really is a bound companion or not, 
can be checked by spectroscopy (whether the atmosphere is as cool as expected)
or proper motion (whether primary and companion are co-moving).
For some of the HST companion candidates detected also in the ground-based images, 
we do have already several epochs. However, in the ground-based images, we cannot 
measure the separation between the primaries and their faint companion candidates 
as precise as in the HST images (only to $\pm 0.5$ pixels in the ground-based images), 
so that we cannot yet measure their proper motions. We can only say that the 
ground-based separations are not inconsistent with those measured in the HST images.

The typical proper motion of other Cha I T Tauri stars is 
$\mu _{\alpha} \simeq -29$ mas/yr and $\mu _{\delta} \simeq 11$ mas/yr 
(Frink et al. 1998, Teixeira et al. 2000).
The proper motions of Cha H$\alpha$ 1 to 12 are not known, yet, but we
can assume that they are similar to other members, because several other indicators 
show that Cha H$\alpha$ 1 - 12 are members;
e.g. they share the typical radial velocity with other Cha I members 
(Joergens \& Guenther 2001). Given the R- \& I-band HST PC images, we can measure
the separation between primary and companion candidate to $\pm 33$ mas (see below).
Hence, with the HST PC pixel scale (45.5 mas), we can check for common proper motion 
after a few years.

\begin{figure}
\vbox{\psfig{figure=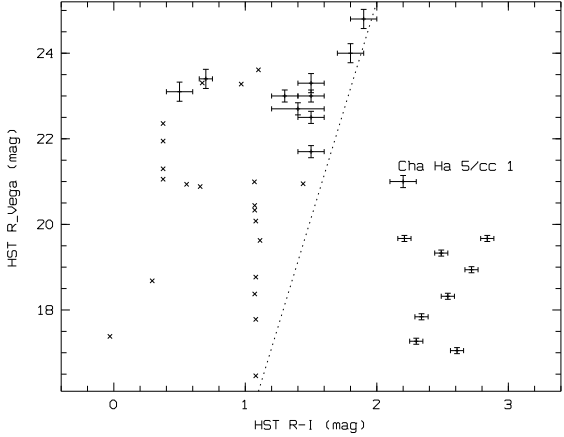,width=10cm,height=7cm,angle=270}}
\caption{Optical color-magnitude diagram with R$_{\rm V}$ versus R$_{\rm V}-$I$_{\rm V}$.
We plot all primary objects (with small error bars), companion candidates 
within 7$^{\prime \prime}$ (with larger error bars), and other objects 
(without error bars) detected in the HST images. 
All data from Table 3. Note the bimodal distribution: All primaries are located 
in the lower right of the figure, together with one companion candidate
(Cha H$\alpha$ 5/cc 1), while all other objects are towards the left.
Possibly, all objects left of the dotted line are background,
while those in the lower right are members of Cha I. Because Cha H$\alpha$ 5/cc 1
is located near the primary objects (and because of its small separation), 
it is the most promising companion candidate.}
\end{figure}

\begin{figure}
\vbox{\psfig{figure=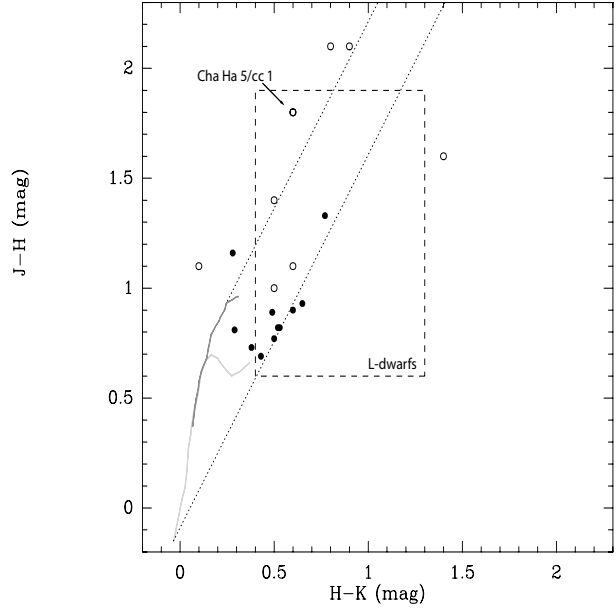,width=8cm,height=8cm}}
\caption{IR color-color magnitude diagram of primaries (filled circles) and companion 
candidates (open circles): J$-$H versus H$-$K$_{\rm s}$ with data from Table 3.
Error bars are omitted for clarity, errors range from $\sim 0.07$ mag for the brightest
primaries to $\sim 0.15$ mag for the faintest companion candidates.
Solid lines indicate the loci of points corresponding to unreddened main-sequence and 
giant stars (Bessel \& Brett 1988). The two parallel dotted lines define the reddening band 
for both main-sequence and giant stars taken from Rieke \& Lebofsky (1985).
The dashed box indicates the locii of L-dwarfs according to Burgasser et al. (2001)
and Leggett et al. (2002).
The companion candidate located to the right of both dotted lines is Cha H$\alpha$ 8/cc 7.
The lefternmost companion candidate (bluest in H$-$K$_{\rm s}$) is 
Cha H$\alpha$ 4/cc 2, located near the early T-dwarf area.
The most promising companion candidate (Cha H$\alpha$ 5/cc 1) is indicated by an arrow.}
\end{figure}

We compiled the R \& I magnitudes of all objects detected in the seven HST images.
From the number of objects brighter than a particular companion candidate
(Fig. 6), we can estimate the background probability.

Lets discuss as example the close companion candidate near Cha H$\alpha$ 5:
There are in total 0.32 (0.23) objects per square arc minute
detected in R (and I, respectively) with magnitude brighter than
Cha H$\alpha$ 5/cc 1 (R=21.2 and I=19.2), see Fig. 6 for the
$\log$~N-$\log$~S of all detected HST sources (primaries, companion
candidates, and other certainly unrelated objects).
Hence, the probability to find one such faint object within $1.5^{\prime \prime}$ 
around at least one target in an ensemble of eleven Cha H$\alpha$ primaries 
observed is $0.69\%$ in R (and $0.49\%$ in I), i.e. quite small.
See also Fig. 8, where it is shown that no other faint objects
are detected in the whole PC field-of-view around Cha H$\alpha$~5.

\begin{figure}
\vbox{\psfig{figure=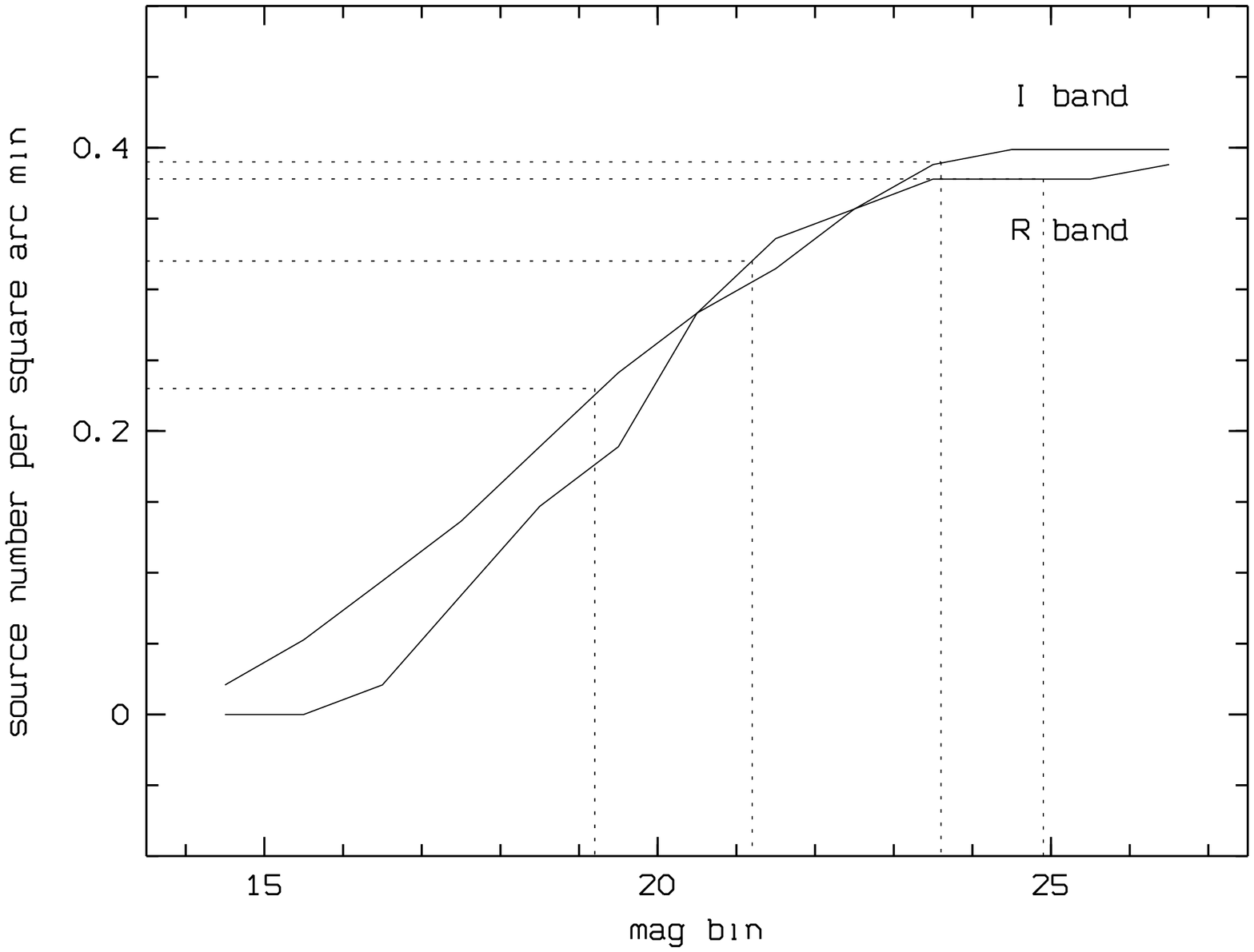,width=10cm,height=7cm,angle=270}}
\caption{Number of sources brighter than limiting magnitude per
square arc minute versus magnitude for R- and I-band HST images.
We used these $\log$~N-$\log$~S curves for estimating the
probability that a certain companion candidate actually is an
unrelated background object behind the Cha I dark cloud. The broken
lines show those data for the brightest (Cha H$\alpha$ 5/cc 1) and
faintest (Cha H$\alpha$ 10/cc 2 in R and 8/cc 2 in I) companion 
candidates.}
\end{figure}

In the same way, we estimated the background probabilities for
all other companion candidates listed in Table 3.
They range from the values mentioned above ($\le 1~\%$) for the brightest 
companion candidate (Cha H$\alpha$ 5/cc 1 at 1.5$^{\prime \prime}$ separation)
up to $\sim 18~\%$ (at 7$^{\prime \prime}$ separation)
for the faintest detected companion candidates,
Cha H$\alpha$ 10/cc 2 and Cha H$\alpha$ 8/cc 2.
Given the whole ensemble with eleven primaries observed
with HST and 16 detected companion candidates, 
we can expect a few real companions.

Following the methodology of Lada et al. (1994) we computed the
extiction map of the target field making use of the H and K$_{\rm s}$
band images taken with SofI. The known foreground stars and embedded
cloud members (see C1999 and C2000) were not used in computing this
map. Because there were no observations of the unreddened background
stellar field (necessary for the determination of the zero point of
the extinction scale), we adopted a $<$H$-$K$_{\rm s}>$
color for this background. Regarding this choice, Alves et al. (1998)
found $<$H$-$K$_{\rm s}> = 0.20 \pm 0.13$ mag towards a complicated
region in Cygnus, essentially at the Galactic plane, while Alves et
al. (2001) found $<$H$-$K$_{\rm s}> = 0.12\pm0.08$ mag towards a
{\em clean} line-of-sight towards the Galactic bulge, 
at $b \simeq 7^{\circ}$. The stellar background towards the Chamaeleon complex 
is probably even better behaved, because it lies at $b \simeq 15^{\circ}$. 
On the other hand, the $<$H$-$K$_{\rm s}$$>$ color of the North Galactic
pole from 2MASS data is also $\sim 0.12$ mag (M. Lombardi, private
communication), so we decided to adopt this as the mean Chamaeleon
background color. We also adopt a conservative intrinsic dispersion in
this measurement of 0.13 mag, i.e. an intrinsic extinction
measurement rms of A$_{\rm V} \simeq 2$ mag.

We show our map in Fig. 7 for the area around Cha H$\alpha$ 4, 5, 8,
10, and 11, where companion candidates were detected.  The spatial
resolution of the map is 50$^{\prime \prime}$ and the extinction
A$_{\rm V}$ ranges from 4 to 22 mag. Each position of this map is
associated with a total extinction which includes the whole Cha I dark
cloud and its fore- and background.

C2000 derived the individual reddening by comparing the observed
colors to intrinsic colors for the known spectral types.
Those absorptions range from A$_{\rm V} \simeq 0.0$ to 1.0 mag (C2000)
and are smaller than the extinction values towards their immediate
background, as measured in the extinction map.
The total extinction towards the immediate background of the primaries 
shown in Fig. 7 range from A$_{\rm V}\simeq $ 6 to 8 mag.
Hence, the primaries are most certainly located
near the front edge of the dark cloud.

\begin{figure}
\vbox{\psfig{figure=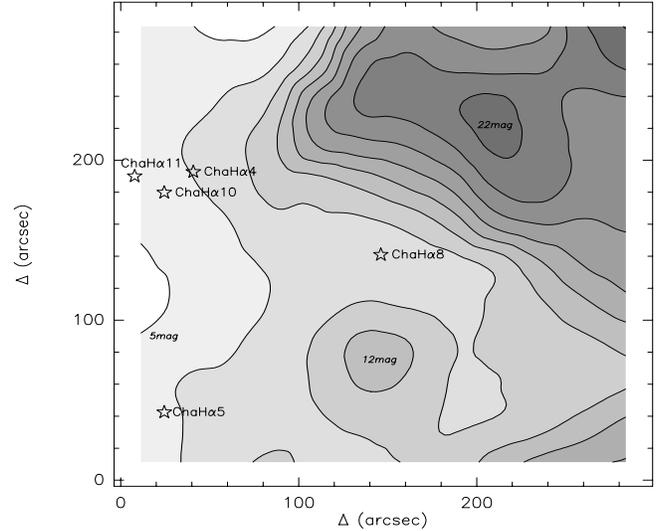,width=8.5cm,height=7cm,angle=270}}
\caption{Background map:
Dust extinction map of the studied region as derived from the IR
(H$-$K) observations of background sources.  Contours start at 5
visual mag of extinction and increase in steps of 2 mag up to 21 mag
of visual extinction.  The effective resolution of the map is 50$^{\prime \prime}$
($\sim 0.04$ pc at the distance of Cha I). The positions of Cha
H$\alpha$ 4, 5, 8, 10, and 11, where companion candidates were
detected, are also indicated.}
\end{figure}

\section{The promising candidate near Cha H$\alpha$ 5}

Cha H$\alpha$~5 is an M6 dwarf with H$\alpha$ and X-ray emission as well as Lithium 
absorption; no radial velocity variations were detected in UVES spectra, consistent 
with membership to the Cha I association (C1999, C2000; Joergens \& Guenther 2001).
Based on a comparison of its location in the H-R diagram with different
tracks and isochrones, Cha H$\alpha$~5 has a mass of
$\sim 0.10$~M$_{\odot}$ and an age of $\sim 2$ Myrs, i.e.
co-eval with the other M6- to M8-type objects Cha H$\alpha$~1 to 12 (C2000).
Hence, it is either a very low-mass T~Tauri star or a relatively massive brown dwarf.

A faint companion candidate is located
$1.085 \pm 0.027^{\prime \prime}$ S
and $1.036 \pm 0.028^{\prime \prime}$ W of Cha H$\alpha$~5,
i.e. at a separation of $1.500 \pm 0.033^{\prime \prime}$
with a position angle of $223.66 \pm 0.10^{\circ}$
(taking into account an uncertainty of $\pm 0.08^{\circ}$ in the N-S
alignment of the detector), identical in both the R- and I-band image.
The companion candidate is not detected in H$\alpha$, while all primary
objects including Cha H$\alpha$~5 are detected.
The HST R- and I-band images of Cha H$\alpha$~5 with its faint
companion candidate are shown in Fig. 8. 
The companion candidate is also marginally resolved 
in the ground-based VRIJHK$_{\rm s}$ images (Figs. 9 and 10).
In the ground-based image, where it is resolved best, the SofI K$_{\rm s}$-band image
from March 2001, the separation measured is $1.38 \pm 0.21^{\prime \prime}$
at a position angle of $220.6 \pm 8.7^{\circ}$,
i.e. much less precise than the HST data, but not inconsistent 
with low differential proper motion.

In the H- and K$_{\rm s}$-band images (Fig. 10), we could obtain the 
magnitude of the companion candidate by normal aperture photometry.
However, in the VRIJ images, this was not possible, because
of the even higher dynamic range.

In the V- and J-band images, we first obtained the flux in each pixel
within 4$^{\prime \prime}$ of the center of the primary, then plotted the
flux as a function of the separation to the primary (Fig. 9, right panels).
Those functions are a superposition of two Moffat functions (Moffat 1969),
one for the bright primary and one for the faint companion candidate
located in the wing of the primary's PSF, both with the same FWHM.
The faint companion candidate
is detected as a small bump in the primary's PSF.
Then, we derived the peak flux of the companion candidate at 1.5$^{\prime \prime}$
separation from the primary by subtracting the flux value of the
upper envelope by the flux value of the lower envelope, both at 1.5$^{\prime \prime}$.
From this flux and the peak flux of the primary, we derived
the magnitude difference between primary and companion candidate.
The same procedure was performed for the VLT R- and I-band images,
where the companion candidate is also detected in the PSF wing.
The magnitudes derived from these VLT R- and I-band images are consistent with
the HST data, where the companion candidate is well detected and well resolved
from the primary ($\pm 0.10$ mag in R and $\pm 0.19$ mag in I).
Hence, we can trust our procedure.

\begin{figure}
\vbox{\psfig{figure=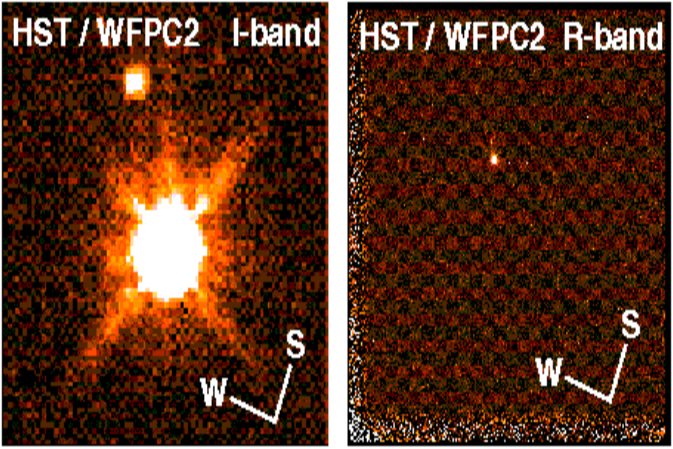,width=9cm,height=6cm}}
\caption{HST R- \& I-band images of Cha H$\alpha$~5 and surrounding field.
Left: I-band image (4$^{\prime \prime} \times 4 ^{\prime \prime}$ field).
A companion candidate 1.5$^{\prime \prime}$ SW of Cha H$\alpha$~5 is detected.
Right: R-band image with the whole PC field
($\sim 15^{\prime \prime} \times 15^{\prime \prime}$), where no objects
other than the primary and one companion candidate are detected.
Also in I and H$\alpha$, no other objects are detected.}
\end{figure}

\begin{figure}
\vbox{\psfig{figure=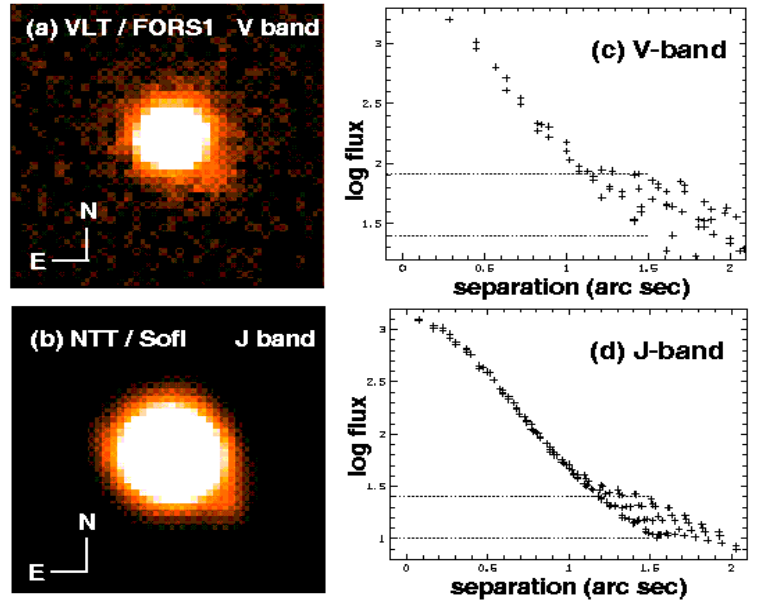,width=10cm,height=8cm}}
\caption{V- \& J-band images of Cha H$\alpha$~5
and its companion candidate 1.5$^{\prime \prime}$ to the SW.
{\bf a.} VLT V-band image. {\bf b.} NTT J-band image.
{\bf c} and {\bf d.} Cha H$\alpha$~5 PSF with the faint companion
candidate as bump in the primary's wing, plotted as
log of flux versus separation in arc sec from the primary,
in V and J, respectively. We computed the companion's
magnitude by subtracting the contribution of the primary
at 1.5$^{\prime \prime}$ separation (lower envelope
of the plotted points) from the peak of the bump 
(upper envelope) and comparing it to the
observed flux and known magnitude of the primary.
The upper horizontal dotted lines show the maximum flux of the companion
in V and J, respectively, and the lower dotted lines show the flux 
contribution of the primary at 1.5$^{\prime \prime}$ separation in V
and J, respectively.}
\end{figure}

\begin{figure}
\vbox{\psfig{figure=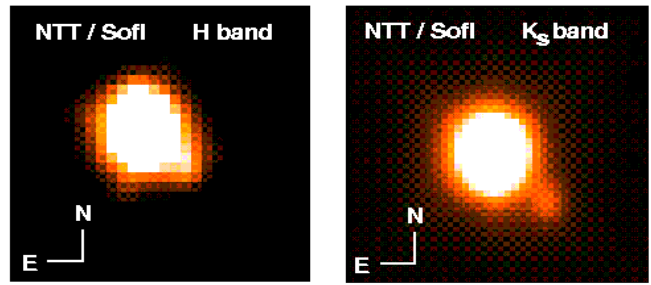,width=8.6cm,height=3.8cm}}
\caption{NTT H- \& K$_{\rm s}$-band images of Cha H$\alpha$~5.}
\end{figure}

For the line-of-sight absorption in the direction of Cha H$\alpha$~5,
we measure A$_{\rm V} \simeq 6.5$ mag (see Fig. 7), i.e. several times more than
foreground to Cha H$\alpha$~5 (A$_{\rm V}$ = 0.98 mag, C2000).
Appart from the fact that this object could be a true companion,
we have to consider several alternatives:
\begin{itemize}
\item The VHK$_{\rm s}$ colors of the companion candidate are consistent 
with a K giant reddened by A$_{\rm V} \simeq 6.5$ mag, which would
be located at $\sim 11$ kpc, i.e. $\sim 3$ kpc from the galactic plane, 
but its dereddened J$-$H color is $1.5~\sigma$ redder than a K giant.
\item The unreddened optical colors of the companion candidate are consistent 
with those of an M5-6 dwarf, which would lie foreground to the extinction.
However, the H \& K$_{\rm s}$ flux would be too bright and the V$-$H and 
V$-$K$_{\rm s}$ colors would be $4 \sigma$ redder than a M5-6 dwarf. 
The bright H and K$_{\rm s}$ fluxes could be IR excess
due to circumstellar material, but such strong excess ($\sim 1.5$ mag
in H and K$_{\rm s}$) was never observed for a mid-M type star.
\item If the object would be a moderately extincted 
(A$_{\rm V} \simeq 1$ mag) late M-dwarf in the background, 
its optical colors would be $2~\sigma$ too blue for a late M-dwarf.
\end{itemize}
Hence, the companion candidate is
unlikely to be a background giant or a foreground dwarf.
As discussed in the previous section, a background object is very unlikely 
anyway. Even considering the whole ensemble observed, the probability to 
find one such object within $1.5 ^{\prime \prime}$ of one our targets 
is $\le 0.7~\%$. The probability for the object being a foreground dwarf 
will be considered below.

For a companion to Cha H$\alpha$~5, we can assume the same extinction as
towards the primary (A$_{\rm I}$ = 0.47 mag, C2000).
The error should be $\sim 0.07$ mag,
as estimated from the errors in R and I.
With the Rieke \& Lebofsky (1985) extinction law, we can then
estimate dereddened colors (Table 5), to be compared with
intrinsic colors of late-M and L dwarfs
(see e.g. Kirkpatrick et al. 1999 \& 2000, henceforth K1999 and K2000).
Few V-band magnitudes of L-dwarfs are available so far,
so that we could not compare our object to typical L-dwarfs in this regard.
The comparison of RIJHK$_{\rm s}$ colors in Table 5 shows
that a spectral type of early- to mid-L is most likely for our object.
The observed magnitude difference between the M6 primary and its companion of
3.8 to 4.7 mag and also its non-detection in H$\alpha$ are consistent with
spectral type L (K1999, K2000).
As seen in Table 5, neither all the observed nor all the dereddened colors are
perfectly consistent with any single spectral type sub-class, but this is not surprising,
because all (or most) known L-dwarfs are much older and more massive,
and evolutionary effects in the atmosphere are expected.
The RIJK$_{\rm s}$ colors of our companion candidate are consistent with
those of young early-L dwarfs in $\sigma$ Ori (Bejar et al. 1999).

\begin{table}
\caption{Colors of Cha H$\alpha$ 5/cc1 and typical L dwarfs}
\vspace{-.3cm}
\begin{tabular}{lcccll} \hline
              & dereddened   & mean  & best  & possible & ref. \\
              & color [mag]  & L0    & guess & range    &      \\ \hline
(R$-$I)$_{0}$           & $1.46 \pm 0.25$ & 2.0 & M7 & M6 - L2   & 5    \\
(R$-$K$_{\rm s}$)$_{0}$ & $5.93 \pm 0.33$ & 6.0 & M8 & M6 - L2   & 5    \\
(I$-$J)$_{0}$           & $2.21 \pm 0.28$ & 2.5 & M8 & M8 - L4   & 1,3  \\
(I$-$H)$_{0}$           & $3.90 \pm 0.28$ & 3.5 & L1 & M9.5-L1.5 & 5    \\
(I$-$K$_{\rm s}$)$_{0}$ & $4.47 \pm 0.34$ & 4.2 & L0 & M9 - L4   & 1,5  \\
(J$-$H)$_{0}$           & $1.69 \pm 0.29$ & 1.0 & L5 & L4 - L8   & 2-4  \\
(J$-$K$_{\rm s}$)$_{0}$ & $2.26 \pm 0.34$ & 1.4 & L5 & L2 - L8   & 1-4  \\
(H$-$K$_{\rm s}$)$_{0}$ & $0.56 \pm 0.35$ & 0.6 & L1 & L0 - L5.5 & 2,4,5\\ \hline
\end{tabular}
Ref.: (1) K1999, (2) K2000
(3) Leggett et al. 2002, (4) Burgasser et al. 2001,
(5) astro.nmsu.edu/\~{}crom/bdwarfs
\end{table}

The Cha H$\alpha$ 5 companion candidate could also be an isolated young
L dwarf in the Cha~I association, rather than a companion of Cha H$\alpha$~5.
This is, however, very unlikely: Even if there are 10 times more L-dwarfs
in Cha I than M-dwarfs (there are 19 M-dwarfs known in Cha I), 
then we would expect to find in total 0.01 L-dwarfs within
$1.5^{\prime \prime}$ around the eleven Cha H$\alpha$ objects observed.

There is an L2.5 dwarf with similar color\footnote{2MASS091838.2+213406
with I~$\simeq 18.68$, K$_{\rm s} = 4.21 \pm 0.07$, and H$-$K$_{\rm s} = 0.43 \pm 0.09$
mag at $\sim 45$ pc (K1999)},
so our object could be an old foreground L2.5 dwarf.
Reid et al. (1999) expect 4634 L-dwarfs with K$_{\rm s} \le 14.5$ mag
on the whole sky, so the probability to find one L-dwarf within
$1.5^{\prime \prime}$ of the eleven Cha H$\alpha$ objects observed
is $0.007~\%$.

With an estimated spectral type of L, the temperature is
$\sim 2200$ to 1400 K (K1999, K2000, Basri et al. 2000, Burgasser et al. 2001).
For the age, we assume 1 to 5 Myrs, as the age of Cha H$\alpha$~5 according
to Burrows et al. (1997) and Baraffe et al. (1998) models, see C2000.

If we use B.C.$_{\rm I} = 0.59$ mag as given by C2000
for M8 also for our companion candidate,
then we obtain $\log ($L$_{\rm bol}/$L$_{\odot}) = -3.0$.
With B.C.$_{\rm K} = 3.2$ mag as given by Leggett et al. (2002) for L1,
we obtain $\log ($L$_{\rm bol}/$L$_{\odot}) = -2.7$.
Considering the magnitude difference in VRIJHK$_{\rm s}$ between primary
and companion being 3.8 to 4.7 mag and the primary having
$\log ($L$_{\rm bol}/$L$_{\odot}) = -1.31$ (C2000),
the companion has $\log ($L$_{\rm bol}/$L$_{\odot}) = -2.8$ to $-3.2$.
Hence, we will use $\log $(L$_{\rm bol}/$L$_{\odot}) = -3.0 \pm 0.3$.

This luminosity, T$_{\rm eff} \simeq 1400$ to 2200 K, and an age of 1 to 5 Myrs
is consistent with a $\sim 3$ to 15~M$_{\rm jup}$ mass object
according to table 3 and figures 7, 9, and 11 in Burrows et al. (1997).
According to figures 6 and 7 in Chabrier \& Baraffe (2000),
an object with the given L$_{\rm bol}$, T$_{\rm eff}$, and age
would have a mass of $\sim 7$ to 15~M$_{\rm jup}$.
Hence, the faint object could be a planet or a brown dwarf, if really a
companion of Cha H$\alpha$~5.
However, all these models may be uncertain at this very young ages,
because they start after the formation with assumed thermal structures.
Grey dynamical calculations of the formation process locate this candidate
between the tracks of a young 10~M$_{\rm jup}$
and a 1~M$_{\rm sat}$ protoplanet (G. Wuchterl, pers. comm.).

The separation between Cha H$\alpha$ 5 and its possible companion is $1.5^{\prime \prime}$
corresponding to a projected physical separation of 210 AU (at 140 pc), i.e. larger 
than expected for planets, but not larger than typical circumstellar disks.
Such a large separation should not be excluded, 
not even for planetary companions, because of possible dycamical
interaction between proto-planets (by which one gets kicked out) and also,
because some of the radial velocity planet candidates have large eccentricities,
i.e. are located at large separations most of the time.

Confirmation of this candidate as true companion can be done best and soon by
spectroscopy, which is scheduled for the VLT for early 2002. If confirmed as true companion, 
this object may be the lowest-mass companion ever detected by direct imaging.

\section{Is Cha H$\alpha$ 2 a close binary~?}

In the previous sections, we investigated wide companions,
which were clearly resolved, at least in the HST images.
To search for very close companions, we determined possible elongations
of the point-spread-functions (PSF) of the primaries, measured as ratio between
the Full Width at Half-Maximum (FWHM) in the direction of the strongest elongation 
and the FWHM in the direction perpendicular to the elongation.
In Fig. 11, we plot the observed PSF elongation versus the magnitude
for all measurements in R, I, and H$\alpha$. 

While most PSFs are consistent with a single point source (elongation $\sim 1$),
there is one notable outlyer, namely Cha H$\alpha$ 2,
where the PSF for both the R- and H$\alpha$ observation is 
elongated by a factor of $\sim 1.5$ towards the NE-SW direction
(not observed in I due to technical problems with WFPC2 shutter).
All three other sources detected on chip WF4 (like Cha H$\alpha$ 2)
as well as Cha H$\alpha$ 9 (on WF2 chip in the same observation)
are consistent with point sources.
The Cha H$\alpha$ 2 PSFs are $\sim 5.5 \sigma$ deviant from the
other Cha H$\alpha$ primaries.
Hence, we do have an indication for Cha H$\alpha$ 2 being a close binary.

Cha H$\alpha$ 2 is elongated in the same direction in both R and H$\alpha$.
Investigating the R-band image of Cha H$\alpha$ 2 in more detail shows
that the SW component is a factor of $\sim 1.56$ brighter than the NE component,
where the flux of the fainter component still includes the flux of the 
brighter one at that separation.
After subtracting the PSF of the brighter (SW) component,
the NE component is fainter by $\sim 0.2$ mag.
The position angle of the NE companion candidate relative to the SW primary 
is $\sim 40 ^{\circ}$ (measured as usually from N to E), and the separation 
between the close pair is almost $0.2^{\prime \prime}$, i.e. slightly less than 
two pixels. As unresolved object, Cha H$\alpha$ 2 has R$_{\rm V}=17.42$ mag (Table 3),
so that the two components would have $\sim 17.45$ and $\sim 17.65$ mag in R$_{\rm V}$.
The probability to find an R$_{\rm V} \simeq 17.65$ mag object with
$\sim 0.2^{\prime \prime}$ separation by chance around 11 primaries observed
is only $3.6 \cdot 10^{-3}~\%$ (obtained from Fig. 6).

\begin{figure}
\vbox{\psfig{figure=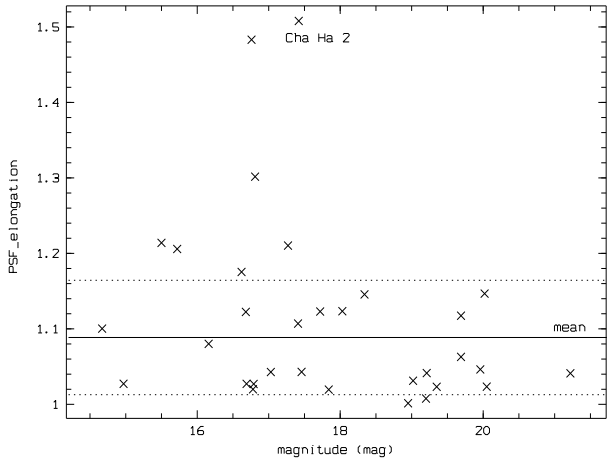,width=9cm,height=6.5cm,angle=270}}
\caption{Apparent elongation of the point-spread-functions (PSF) 
versus magnitude for the primaries (and the Cha H$\alpha$ 5 companion 
candidate). Only Cha H$\alpha$ 2 (observed in R and H$\alpha$, not in I)
appears to be elongated.
All other PSFs have a mean elongation of $1.089 \pm 0.076$ plotted as full 
line (with broken lines indicating its $1 \sigma$ standard deviation range).
The Cha H$\alpha$ 2 PSFs with 1.51 in R (and 1.48 in H$\alpha$),
that is $\sim 5.5~\sigma$ deviant from the mean,
hence possibly indicating a very close companion.}
\end{figure}

Even though the PSF elongation of Cha H$\alpha$ 2 is cleary larger 
than in all other objects, we regard it as candidate binary.
See Fig. 12 for the HST image of this object.
We cannot completely exclude that the apparent elongation could
be due to, e.g., bad pixels.
Further obvervations are needed to confirm the apparent elongation.
The ground-based FORS1 and SofI images do not have sufficient resolution
to detect this elongation.

\begin{figure}
\caption{The possible marginal $\sim 0.2^{\prime \prime}$ elongation of Cha H$\alpha$ 2:
(a) HST WF4 chip R-band image of Cha H$\alpha$ 2.
(b) 3D representation of the PSF of Cha H$\alpha$ 2 (R-band)
showing a faint secondary bump to the NE.
(c) HST WF4 chip image of Cha H$\alpha$ 2 in H$\alpha$, again slightly elongated.
(d) 2D representation of the PSF of Cha H$\alpha$ 2 (R-band)
showing the small SW-NE elongation.
(e) HST WF4 chip R-band image of another star (same chip, same observation).
(f) PSF of that other star.
While the PSF of Cha H$\alpha$ 2 is always slightly elongated,
the PSF of the (anonymous) comparison star is consistent with a point source.
{\bf See A\&A paper for figure.}}
\end{figure}

Cha H$\alpha$ 2 is located close to the sub-stellar limit,
i.e. either a very low-mass star or a relatively massive brown dwarf,
according to both Burrows et al. (1997) and Baraffe et al. (1998) tracks, see C2000.
As approximately equal-brightness
and, hence equal-mass binary, both components would still be very
close to the sub-stellar limit, i.e. either low-mass stars or brown dwarfs.
More precise imaging photometry (e.g. with the PC or with NICMOS)
and resolved spectroscopy may resolve this issue.
A separation of only $\sim 0.2^{\prime \prime}$ would correspond to
only $\sim 28$~AU at 140 pc, so that one could hope to see orbital motion
within a few decades.

\section{Discussion}

From the images of Cha H$\alpha$ 5 and 8 and their companion candidates
observed on the PC chip, we determined the dynamic range achieved on the PC chip.
From the deepest exposures with Cha H$\alpha$ 4, 10, \& 11 on WF4 and Cha H$\alpha$ 8
on WF2, we determined the dynamic range achieved on the WF chips.
In Fig. 13, the dynamic range is plotted as flux ratio versus separation,
with flux ratio being the ratio between either a companion or the $3 \sigma$ background 
level and the particular primary (I-band).

From the mean (and faintest, respectively) I-band magnitude of the primaries
(being 16.0 mag and 17.5, respectively) and the dynamic range limit (log flux ratio being 4, 
i.e. a magnitude difference of 10 mag outside of 2$^{\prime \prime}$ with the PC), 
we can then obtain the magnitude limit for detectable companions,
namely 26 mag (27.5 mag), or one mag brighter at 1$^{\prime \prime}$ separation
(and 2 mag brighter for the WF chips).
For an assumed I$-$K$_{\rm s}$ color index of $\sim 4.5$ mag for an L- or T- dwarf,
this would correspond to a limit of 21.5 mag (23.0 mag) in K$_{\rm s}$.
This limiting magnitude at a distance of 140 pc and an age of 2 Myrs would
correspond to a limiting companion mass of $\le 1$M$_{\rm jup}$
according to table 1 in Burrows et al. (1997) with B.C.$_{\rm K}$=2 mag
(as for Gl 229 B, Leggett et al. 1999).

Hence, we should have been able to have detected all companions with
mass above $\sim 1$M$_{\rm jup}$ outside of 2$^{\prime \prime}$ (320 AU)
and all companions down to a few M$_{\rm jup}$ at $\sim 100$ AU.
Outside of $0.35^{\prime \prime}$ (50 AU), we should have detected
all companions with masses above $\sim 5$M$_{\rm jup}$ (K$_{\rm s} \simeq 18$ mag).

\begin{figure}
\vbox{\psfig{figure=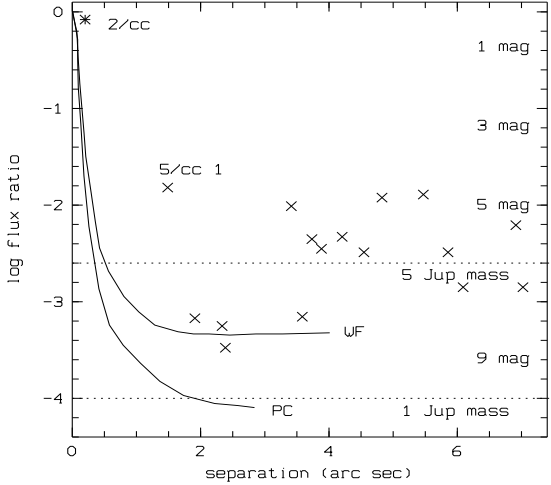,width=12cm,height=7.5cm,angle=270}}
\caption{Dynamic range for our HST WFPC2 images:
We plot the log of the flux ratio between primary and detected companion candidates
(crosses) with I-band magnitudes from Table 3 (magnitude difference is given on
the right-hand side y-axis). Also shown is the I-band dynamic range curve obtained 
on the PC and WF chips (determined as flux ratio between primary and flux being
$3 \sigma$ above the background level). The two dotted lines show the approximate
flux ratio for 5 and 1 M$_{\rm jup}$ mass companions (at 2 Myrs and 140 pc) around 
a primary (calculated for the mean primary I-band magnitude of 16 mag) according to 
Burrows et al. (1997) models. We indicate the locations of the 
very close binary candidate Cha H$\alpha$ 2 (as star symbol in the upper left) 
and the most promising resolved companion candidate Cha H$\alpha$ 5/cc 1.}
\end{figure}

If isolated brown dwarfs are ejected early in the accretion phase (Reipurth \& Clarke 2001)
or some time during the pre-main sequence evolution (Sterzik et al. 2001), one should 
expect more high-mass-ratio binaries 
among very young stars than among old stars, which we can check with our sample.
We observed 11 M6-M8-type primaries with HST and found 16 clearly resolved wide
companion candidates around five of them plus one additional very close 
($0.2^{\prime \prime}$) candidate binary. From the background population,
we estimated the probability for each candidate to be a true companion,
so that we can expect a few real companions among the candidates.
From the optical and IR colors, only two of the candidates could be
L- or T-type objects (Cha H$\alpha$ 5/cc 1 and Cha H$\alpha$ 4/cc 2). 
Up to two true sub-stellar companions around 11 primaries
correspond to a percentage of $18 \pm 13~\%$.
All these detected companion candidates (as well as un-detected, but detectable companions)
would have separations in the range from $\sim 100$~AU (resolution limit) to 1000~AU 
(somewhat arbitrary upper limit) and high mass-ratios of 10 to 100
(with the ratio being the mass of the primary devided by mass of the companion).
If Cha H$\alpha$ 2 is a close binary (Sect. 7),
then the secondary could also be a sub-stellar companion,
so that we would then have $\le 3$ sub-stellar companions
around 11 primaries, i.e. $27 \pm 16~\%$.

This should be compared to the frequency and orbit characteristics of other 
high-mass-ratio binaries with large separations, i.e. brown dwarfs in orbit around a star.
Recently, Gizis et al. (2001) estimated the frequency of wide, visual, 
old L- and T-type brown dwarf companions to normal stars to be $18 \pm 14~\%$.
This number is consistent with our estimate given above, 
so that we find no evidence for the ejection scenario.
However, because of the large error bars due to small-number-statistics,
we do not have sufficient evidence to make a strong statement in 
this regard. In addition, 
the surveys discussed by Gizis et al. (2001)
and our survey have different dynamical ranges and detection limits.

We also have to refrain from comparing our results with previous surveys
for companions around young low-mass stars in Chamaeleon 
(Reipurth \& Zinnecker 1993, Brandner et al. 1996, Ghez et al. 1997, 
K\"ohler 2002), because our sample of primaries (M6- to M8-type bona-fide
and candidate brown dwarfs) is different from the previously surveyed samples
(G- to early M-type T~Tauri stars)
and because the previous surveys were restricted to smaller dynamical ranges,
up to a magnitude difference of 5 mag between primary and companion candidate,
while all but one (Cha H$\alpha$ 5) or two (Cha H$\alpha$ 2) of our companion 
candidates have larger magnitude differences.

In our observations, we found one promising companion candidate (Cha H$\alpha$ 5/cc 1),
which could be a brown dwarf (or even giant planet) companion.
The other candidates are all fainter and often detected only in I.
If they were true companions, they would all be giant planets, given their
magnitudes, assuming the same age and distance as towards the primaries.
However, they are all at physical separations $\ge 200$ AU, which is not
expected for planets, but should not be excluded.
Follow-up 2nd epoch imaging and/or spectroscopy will show,
whether and which of our companion candidates are truely cool and bound.

\acknowledgements{We would like to thank the support staff at VLT, NTT, and STScI,
especially our program coordinator Galina Soutchkova, as well as
G\"unther Wuchterl, Eike Guenther, and Eduardo Mart\'\i n for useful discussion.
RN and WB acknowledge the hospitality and support from the Institute for Astronomy 
at the University of Hawai'i, where part of the work was carried out.
RN acknowledges financial support from the BMBF through DLR grant 50 OR 0003.
VJ acknowledges financial support from the DFG Schwerpunktprogramm Sternentstehung.}

{}

\end{document}